\documentclass[submission,copyright,creativecommons]{eptcs}
\providecommand{\event}{FESCA 2016} 
\usepackage{breakurl}             
\usepackage{acronym}
\usepackage{verbatim}
\usepackage{hyperref}
\usepackage{multirow}
\usepackage{xspace}
\usepackage{color}
\usepackage[usenames,dvipsnames]{xcolor}
\usepackage{graphicx}
\usepackage{listings}
\usepackage[T1]{fontenc}

\newcommand{\ec}{e-commerce}
\newcommand{\etal}{{\it et al.}\xspace}

\acrodef{ACE}{Attempto Controlled English}
\acrodef{OWL}{W3C Web Ontology Language}
\acrodef{NL}{Natural Language}
\acrodef{RDF}{Resource Description Framework}
\acrodef{CNL}{Controlled Natural Language}
\acrodef{SPARQL}{Simple Protocol and RDF Query Language}
\acrodef{uQL}{Use Cases Query Language}
\acrodef{uCat}{Use Cases Analysis Tool}
\acrodef{RUS}{Restricted Use Case Statements}
\acrodef{SQL}{Structured Query Language}
\acrodef{EUC}{Essential Use Cases}
\acrodef{RUST}{Restricted Use Case Statements Template}
\acrodef{SCARP}{SCenario bAsed Rapid software Prototype}
\acrodef{UML}{Unified Modeling Language}
\acrodef{OMG}{Object Management Group}
\acrodef{uCat}{Use Cases Analysis Tool}
\acrodef{SWRL}{Semantic Web Rule Language}
\acrodef{SUS}{System Usability Scale}

\definecolor{light-gray}{gray}{0.98}
\definecolor{dark-blue}{rgb}{0,0,0.5}

\newcommand{\thetitle}{Validating an Approach to Formalize Use Cases with Ontologies}
\title{\thetitle}

\author{Rui Couto \and Ant\'{o}nio Nestor Ribeiro \and Jos\'{e} Creissac Campos
\institute{HASLab/INESCT TEC \& Dept. of Informatics/University of Minho\\Portugal\footnote{This work is financed by the ERDF - European Regional Development Fund through the Operational Programme for Competitiveness and Internationalisation - COMPETE 2020 Programme  
within project <<POCI-01-0145-FEDER-006961>>, and by National Funds through the FCT - Fundação para a Ciência e a Tecnologia (Portuguese Foundation for Science and Technology) as part 
of project  UID/EEA/50014/2013.}}
\email{\{rui.couto, anr, jose.campos\}@di.uminho.pt}
}
\def\titlerunning{\thetitle}
\def\authorrunning{R. Couto, A. N. Ribeiro \& J. C. Campos}
\begin{document}
\maketitle


\begin{abstract}
Use case driven development methodologies put use cases at the center of the software development process. However, in order to support automated development and analysis, use cases need to be appropriately formalized. This will also help guarantee consistency between requirements specifications and developed solutions. Formal methods tend to suffer from take up issues, as they are usually hard to accept by industry. In this context, it is relevant not only to produce languages and approaches to support formalization, but also to perform their validation. In previous works we have developed an approach to formalize use cases resorting to ontologies. In this paper we present the validation of one such approach. Through a three stage study, we evaluate the acceptance of the language and supporting tool. The first stage focusses on the acceptance of the process and language, the second on the support the tool provides to the process, and finally the third one on the tool's usability aspects. Results show test subjects found the approach feasible and useful and the tool easy to use.
\end{abstract}

\section{Introduction}

Requirements are the foundation for the software development process.
It is thus somewhat paradoxical that software development suffers greatly from inconsistencies between requirements and the final resulting solutions.
Indeed, such inconsistencies are the major root cause of project failures~\cite{kamata2007does}.
Creating and updating requirements models is a laborious task. As the software development process goes on, the focus naturally leans towards the development process, and such models end up being put aside. 
Moving the focus from modeling to development makes it harder to guaranteeing the compliance between requirements and the final software solutions.
Thus, compliance between requirements and software solutions is still one of requirements engineering concerns \cite{DBLP:conf/wicsa/CaraccioloLN15}. 
As requirements tend to be validated only at the end of the development process, fixing possible inconsistencies will have a big impact in the development process and in the software solution. The further the final solutions are from the requirements (models), the higher is the probability for the software solutions to fail.

Use cases~\cite{Jacobson:2004:OSE:993806} are a relevant tool for requirements specification. Use case models provide two levels of abstraction: graphical diagram (\emph{Use Case Diagrams}), and, textual descriptions (\emph{Use Case scenarios}).
By using a clear diagram notation and simple textual formats they are a good bridge between developers and stakeholders. 
Use case driven development methodologies (e.g. \cite{Rosenberg:2013:UCD:2440179, DBLP:conf/seke/SubramaniamLFE04}) put use cases at the center of the software development process in order to deal with inconsistency issues. Those methodologies focus on use cases in order to guide the development process. 
In doing so, however, they have to face some challenges.
Since use case scenarios are described in textual formats, they are hard to analyze. Manual analysis is a laborious process, vulnerable to subjectivity. Tools that help solving these issues by automatically extracting information from use cases to guide the development process are required. 

Existing tools tend to focus on specific aspects of software development, hence not providing a systematic approach to development.
For instance, some tools generate early prototypes (e.g.~\cite{DBLP:conf/somet/KamalrudinAGHR14}), while others focus on the testing phases (e.g. \cite{nebut2006automatic}). 
Additionally, not all tools address use cases formalization in a systematic way. 
On the one hand, the lack of systematic approaches makes the extraction processes more prone to subjectivity in the analysis.
On the other hand, improper formalization approaches lead to  harder to maintain and update representations.

In order to support the analysis of use case specifications, we propose a tool-supported approach that enables representing the information use cases contain as a knowledge base.
The knowledge base is automatically created from the specifications, by the uCat tool \cite{DBLP:journals/corr/CoutoRC14}. 
Contrarily to textual formats, formalized specifications (as a knowledge base is) are easier to maintain, while providing means to query and analyze their information. The formalization also removes the ambiguity resulting from natural language. 
By providing means for early analyze and validation of the requirements it is possible to reduce the possibility for inconsistencies in the final solutions.
The knowledge base defines the basis for the automation of these analysis activities. Examples of automated activities are requirements patterns inference, formal verification (e.g. model checking), or test case generation.
The process requires
a language to specify the use cases in human-readable form and another to express the knowledge base.
For the former we resort to a \ac{CNL} \cite{DBLP:journals/corr/CoutoRC14}, for the latter to
an ontology language, namely the \ac{OWL} \cite{owl:site}.

The usefulness of formal methods tends to be hindered by resistance to their adoption. Indeed, despite the value provided by formal approaches, it is also necessary to consider the cost implied in having to learn and apply them. Hence, when considering formal engineering approaches, it is not only relevant to propose new languages and tools, but also to validate their usefulness and acceptance by the target users. This is the main focus of this paper. In previous work we addressed the formalization of use cases~\cite{DBLP:journals/corr/CoutoRC14}. Our objective now is to validate the users's acceptances of the approach and measure how willing users are to accept the formalization technique that we are developing.

In this paper we focus on the validation of the \ac{CNL} required to handle the use case specifications, as well as the tool supporting it. The study analyzes how the language performs in supporting the user statements, and how the tool supports the language. The study analyses also the advantages provided by the tool, regarding the process. 
The remainder of this paper is structured as follows. Section~\ref{sect:related} presents  related work,  Section \ref{sect:fa}  presents the formalization approach, and Section \ref{sect:val_ap} the corresponding validation study. Section \ref{sect:disc} presents a discussion of   results. Conclusions and future work are presented in Section~\ref{sect:conc}.

\section{Related work}
\label{sect:related}

This section covers work on requirements specification with use cases. Regarding how use cases can be specified, and how they can be used to drive the development process.

\subsection{Use cases}

Use cases are a standard for capturing the functional requirements of software systems.
No strict template for describing individual use cases has been defined, so different authors propose different approaches. 
It should also be noted that use cases can be used at different stages of requirements analysis and specifications, and with different flavors; from more textual description amenable for analysis, to more operational descriptions useful for specification.  
Tabular representations have proven popular (e.g. as proposed in \cite{Fowler:2003:UDB:861282}). 
This style of representation is based in an operational actor/system interaction style, where the interactions are described sequentially, and is adopted herein. 

A number of authors have explored languages to express interactions.
Some aim at providing powerful, yet complex, languages for use case specification.
An example is the RUCM approach~\cite{Yue:2010:ADU:2049376}, which provides a language with support (for instance) for conditional structures (c.f. \texttt{if}) and other elaborated statements.
On the contrary, \ac{EUC}~\cite{Constantine:1999:SUP:301248} is an approach that succeeds in simplifying use cases in order to provide simpler and more readable statements. 
The author resorts to minimalist statements to create simpler specifications without losing expressiveness.
This kind of specification (while not automable) goes toward the kind of statements we aim for: simple, yet expressive. 

It is also possible to find works which address the extraction of architectural information from use cases.
Yue \etal \cite{yue2011automated} propose an approach to generate class diagrams from use cases. 
Deeptimahanti and Ratna~\cite{DBLP:conf/indiaSE/DeeptimahantiS11}, and, Mala and Uma \cite{DBLP:conf/pricai/MalaU06} address the generation of UML structural diagrams from natural language descriptions. Both approaches are semi-automated, and rely on natural language analysis. 
While these works do not focus on use case formalization, they prove the viability of extracting architectural information from them. However, the analysis of natural language is challenging and authors acknowledge the high variability of the results.

We focus on the direct translation of use cases into OWL. 
To avoid the problems of natural language processing, we aim for a formalized specification language. 
In order to keep it simple, we adopt a tabular format where the complexity of the scenarios conditional behavior is partially hidden by alternative scenarios (meaning approaches such as RUCM unnecessarily complex for our purposes).
With the above in mind, we propose an approach that a) automatically formalizes use cases' information, b) is supported by a simplified, readable yet computable language to describe use cases, and, c) has the expressiveness to support use case specifications. Our approach differs from the existing ones by targeting the formalization of the use cases into a knowledge base, supported by a lightweight language.

\subsection{Use case driven approaches}

Use case driven approaches put use cases at the center of the software development process. Jacobson~\cite{Jacobson:2004:OSE:993806} presents a well known approach where use cases play the central role, supporting the development process from the design phase until testing.
The approach assumes no automation of the development process.
Rosenberg and Stephens~\cite{Rosenberg:2013:UCD:2440179} present the ICONIX process which supports the derivation of object oriented software designs, in the form of \ac{UML} diagrams, from use cases.  
Despite the authors not presenting an automated process, they successfully show how the use cases contain the required information to derive the systems they represent.

Gherkin~\cite{gherkin} is a popular approach, which supports the software development process with scenarios. As with our approach, it also allows the creation of specifications understandable both by developers and stakeholders. Despite scenarios not being use cases, they have fairly similar formats.
The scenarios support the definition of system behaviors, with the purpose of documentation and automated testing. 
Similarly to Gherkin, in our approach we aim to formalize use cases in a simple format. However, we focus on the exploration of the data extracted from use cases rather than in a direct translation to source code as this opens up a wider range of possibilities. 

Subramaniam \etal~\cite{DBLP:conf/seke/SubramaniamLFE04} present the Use Case Driven Development Assistant, a tool to support a use case driven development process. The tool takes a functional description as input, and with a parser generates a use case scenario. By performing textual analysis on the use case, the authors are able to generate class diagrams representing the data therein.
The work shows the viability of resorting to use cases in order to support the software development process. In our work, we put the emphasis in the formalization, as support for the development process. Producing architectural hints is another possibility to use the formalized information.

The works above successfully provide approaches that put use cases at the center of the software development process. Each author presents a different use for use cases, resorting to a custom approach.
By focussing on the formalization of the information in the use cases, we open up the possibility of applying similar use case driven techniques in a more integrated fashion.

\subsection{Usability evaluation approaches}

While evaluating software usability there are several approaches available. Kline and Seffah present an usability study on on IDEs \cite{Kline2005607}. The authors resort to three studies, with four evaluation techniques. From several stated techniques (interviews, task analysis, direct observation, questionnaires, heuristic evaluation) the authors selected interviews, heuristic, psychometric assessment and observation to evaluate a set of IDEs. At the end the authors are able to extract relevant information regarding usability. This work is able to ground the viability of resorting to this kind of techniques in order to evaluate several usability measures, namely regarding the complexity of interfaces. 

Ribeiro \etal resort to PSSUQ and ICF-US usability assessment instruments to evaluate an interactive application~\cite{Ribeiro2015635}. The techniques allowed the authors to identify some impairments of the user interfaces (e.g. difficulties in using the remote control), as well as evaluating several aspects of the interface (e.g. ease of use). This work presents a success use case in the usage of assessment instruments to evaluate several aspects of usability. Another well known tool is the \ac{SUS}~\cite{brooke1996sus}, whose suitability to evaluate IDEs seems also to be appropriate, as demonstrated in this study.

Combining different approaches in order to evaluate our tool seems appropriate, as grounded by the previously presented works. A combination of usability evaluation tools and questionnaire is proposed to evaluate our tool.

\section{The formalization approach}

\begin{figure*}[tb]
	\centering
	\includegraphics[width=0.85\textwidth]{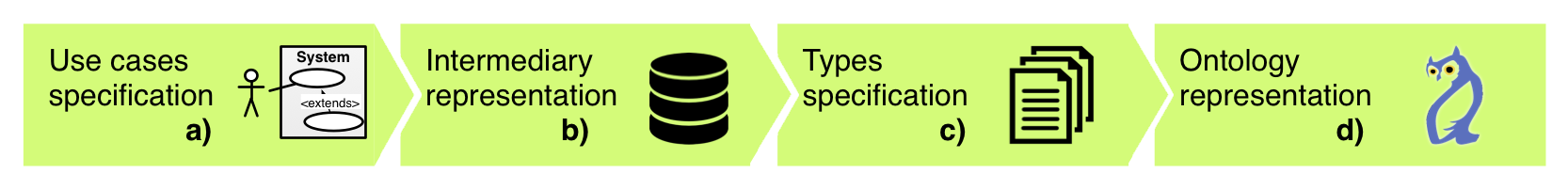}
	\caption{The formalization approach.}
	\label{fig:fa}
\end{figure*}

The formalization approach (c.f. Figure~\ref{fig:fa}) is presented in this section.
Our approach starts with the specification of use cases (\textbf{a)}), which are represented in an intermediary representation (\textbf{b)}). After the provision of additional information (for example, the types for the extracted entities (\textbf{c)}), the methodology is able to produce a knowledge base \textbf{d)}.

\label{sect:fa}

\subsection{Use cases}

In our approach we focus on simple sequential scenarios specifications which cover detailed functional requirements (c.f.~\cite{Fowler:2003:UDB:861282}). Conditional behavior is encoded as alternative scenarios. 
The approach takes as inputs (Figure~\ref{fig:fa}, \textbf{a)}) use case specifications written according to a \ac{CNL}. 
The specification format follows a \emph{User input}/\emph{System response} style, using a tabular representation for simple statements. 

\ac{RUS} is the \ac{CNL} used to create the specifications. \ac{RUS} was inspired in \ac{ACE}, therefore is based also in triples (\emph{subject}, \emph{predicate}, \emph{object}).
\ac{RUS} supports statements such as, for instance, ``\texttt{user selects action}'', ``\texttt{user selects the action}'', or ``\texttt{user selects an action}''. In this case the same data will be extracted (the $<$\texttt{user}, \texttt{selects}, \texttt{action}$>$ triple). \ac{RUS} supports more complex statements such as  ``\texttt{user provides name, email and password}''.
In this case, the extracted data will be the triples $<$\texttt{user, provides, name}$>$, $<$\texttt{user, provides, email}$>$ and $<$\texttt{user, provides, keyword}$>$. By creating a use case scenario resorting to \ac{RUS} it is then possible to automatically achieve a formalized knowledge base.

The RUS language is extensible.
\ac{RUST} is the template mechanism that allows us to define the \ac{RUS} language. \ac{RUST} defines the allowed \ac{RUS} inputs, and how they are mapped into \ac{OWL}. \ac{RUST} contains placeholders (c.f. \texttt{<S> <P> <O>}) that define how each word of the provided \ac{RUS} will be mapped into a triple. In order to extend the language, developers need only to define new \ac{RUST} statements. For the statement ``\texttt{user clicks in the link}'', the corresponding \ac{RUST} is ``\texttt{<S> <P> in the <O>}''. 
In this case, \texttt{<S>} \texttt{<P>} and \texttt{<O>} denote the \emph{subject}, \emph{predicate} and \emph{object}, respectively, while ``\texttt{in the}'' denotes text that should exist in the matching \ac{RUS} statements. 
For \ac{RUS} statements with a variable number of inputs (c.f. ``\texttt{user inserts name, email, password}'' above), it is possible to specify a placeholder (by adding the \texttt{+} symbol) that will handle the several inputs. The corresponding \ac{RUST} is as follows ``\texttt{<S> <P> <O>+}''. 
Further details and a more extensive example of the language can be found in~\cite{DBLP:journals/corr/CoutoRC14}.

Using a \ac{CNL} instead of a formal language (such as \ac{RDF}) has the advantage of improving the readability \cite{DBLP:conf/quatic/CoutoRC14}, while retaining the possibility to automate the processing of the specifications~\cite{DBLP:journals/corr/CoutoRC14}.
By not using free form, \ac{NL}, textual formats we avoid the costs of using textual analysis techniques and deal with their associated challenges (e.g. ambiguity~\cite{DBLP:journals/corr/GhoshELLSS14}). We consider this approach to be a good compromise between expressiveness and tractability, and one which is suitable for our purposes~\cite{DBLP:conf/quatic/CoutoRC14}.

\subsection{Formalized information}

Formalizing use cases according to \ac{RUS} allows us to map them into an intermediary representation (Figure~\ref{fig:fa}, \textbf{b)}). This representation allows users of the approach to review the extracted information, and to provide the types for each extracted entity (Figure~\ref{fig:fa}, \textbf{c)}). This step is required in order to achieve an ontology, as ontology instances (i.e. the extracted information) must belong to a class (i.e. have a type).

From the intermediary representation and the types information, it is possible to achieve an \ac{OWL} ontology (Figure~\ref{fig:fa}, \textbf{d)}), which contains the use cases scenario formalization. 
Both the ontology (structure) and its information (instance) are generated. 
The approach is supported by the \ac{uCat} tool (Figure~\ref{fig:ucat1}), which supports \ac{RUS} specification, and corresponding formalization. The tool, presented in the next section, provides also a reasoner to query the knowledge base, which resorts to the \ac{SPARQL} \cite{sparql} in order to  extract information about the use cases  the from the knowledge base.

\subsection{The uCat tool}
\label{sect:tool}

The \ac{uCat} tool \cite{DBLP:journals/corr/CoutoRC14} supports the automation of the formalization approach, by providing an IDE for use case specification and intermediary information generation.  
\begin{figure}[tb]
\centering
\minipage{0.51\textwidth}
  \includegraphics[width=\linewidth]{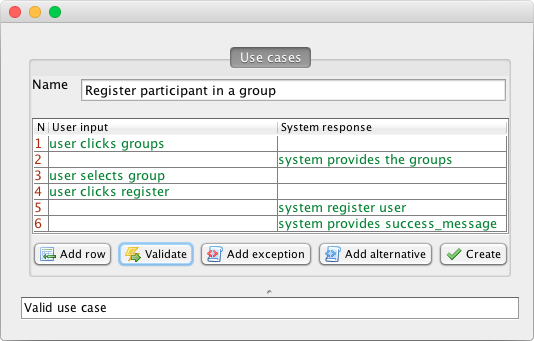}
\endminipage
~
\minipage{0.47\textwidth}
  \includegraphics[width=\linewidth]{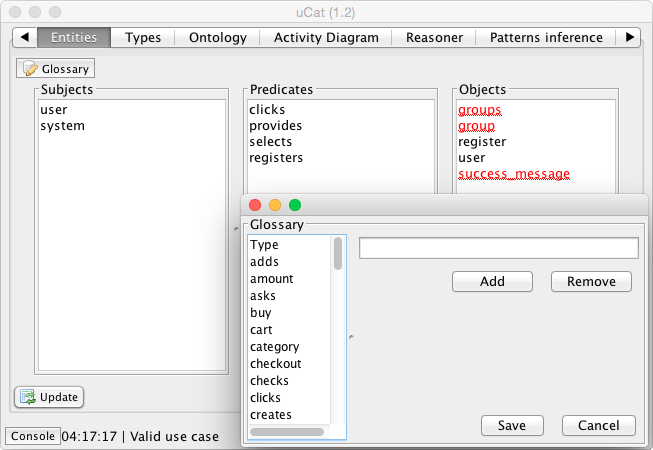}
\endminipage\hfill
  \caption{The uCat tool (left, use case specification interface; right, extracted information).}\label{fig:ucat1}
\end{figure}
\ac{uCat} provides a graphical interface for use cases scenarios input (in \ac{RUS}) (see Figure~\ref{fig:ucat1}, left). In this interface it is possible to add several actors and for each actor several use cases. The interface allows also to check the specifications validity against \ac{RUST}. 
Another feature, entities extraction (see Figure~\ref{fig:ucat1}, right), extracts the subjects, predicates and objects from the use case specifications and supports the definition of types for subjects and objects as explained above. 
At the same time, it supports  a preliminary validation of the extracted information against a glossary that contains the set of terms that are relevant in the context. This goes towards the work of Tena \etal \cite{DBLP:journals/infsof/TenaDDA13}, by providing a vocabulary to produce more standardized specifications.
The tool takes all these inputs and generates an ontology, providing both (raw) textual and graphical representations to preview the extracted information.

\section{Validation of the approach}
\label{sect:val_ap}

While developing our approach several questions arose regarding its viability.  
A first set of questions regarded the \ac{RUS} language. Whether it is expressive enough to handle real world specifications, while easy to use both to write use case and to interpret them. 
A second set of questions regarded the developed tool support (the uCat tool).
The quality of the support it provides to write specifications and, more generally, its overall usability.
To address these issues we carried out a three stage validation of the approach. For each stage, we defined a set of target objectives.

\noindent\textbf{STAGE 1} aimed to validate the expressiveness of the language. 
It addressed three of the main objectives defined for the approach:
\begin{description}
\item \textbf{Objective 1} Provide formalism with a minimal effort for the users;
\item \textbf{Objective 2} Provide a language which is expressive enough to support use case specifications;
\item \textbf{Objective 3} Provide a language which is easy to use.
\end{description}

\noindent In \textbf{STAGE 2} is addressed the tool's support for \ac{RUS}. 
It addressed three more objectives of the approach (in this case, related specifically to uCat):
\begin{description}
\item \textbf{Objective 4} Be easy to learn how to use; 
\item \textbf{Objective 5} Be acceptable as a complement/substitution for other tools;
\item \textbf{Objective 6} Provide a good support for the \ac{RUS} language.
\end{description}

\noindent\textbf{STAGE 3} evaluated the tool's usability, thus addressing the seventh and final objective:
\begin{description}
\item \textbf{Objective 7} Have good usability.
\end{description}

\subsection{Setup of the study} \label{sec:setup}

Due to practical reasons two different groups of test subjects were used. 
A first group of 18 participants (16 male, 2 female), with ages between 20 and 25 years with a mean of 21, performed the first two evaluation stages. They were all Informatics Engineering students at the University of Minho. 
The third stage was performed with a group of 21 participants (18 male, 3 female), all students from a masters course on Software Engineering at the same University. They had an average age of 24 years, and 2.4 years of experience specifying use cases (mostly from academia, some from industry). 
All test subjects from both groups had previous contact with use cases tabular representations, and none had previous contact with \ac{RUS}, the uCat tool or even our work. 
We deliberately selected participants with a software engineering background, as they are the expected users for the tool.

\begin{table}
  \centering
  \caption{Excerpt of a collected use case and its RUS version.} \label{tab:ouc}
  {\scriptsize\tt
   \begin{tabular}{|l|l|l||l|l|l|}
    \multicolumn{3}{l}{The original version} &  \multicolumn{3}{l}{The RUS version}\\
    \hline
    1& inserts project   &                    & 1& user inserts            & \\
     & identifier        &                    &  & project\_id             & \\
    2&                   & confirms project   & 2&                         & system confirms project \\
     &                   & existence          &  &                         & \\
    3& inserts work plan &                    & 3& user inserts work\_plan & \\
    4&                   & confirms insertion & 4&                         & system confirms \\
     &                   & of work plan       &  &                         & insertion \\
    \hline
  \end{tabular}}
\end{table}

For the first two stages, we started by collecting a set of use cases previously produced by the participants as an assignment for a class (unrelated with our study). In total, we selected 8 distinct use case scenarios, that we formalized in \ac{RUS} (see Table~\ref{tab:ouc} for an excerpt of such descriptions). 
This allowed us to obtain a set of \ac{RUS} use cases for a domain known to the test subjects, while avoiding their contact  with the language.
Next, we created  scripts for the participants, containing the tasks for each stage and instructions about how to perform them. A task consists in an exercise the user should perform (e.g. create or interpret a specification). We created also a questionnaire to evaluate the experience of the participants with the language and with the tool  (see Table \ref{tab:quest}). 
Note that questions 1 to 3 and question 16 are open (although a numerical answer was expected), while the remaining questions, up to number 24, were answered in a 7-point Likert scale (0 meaning low and 7 meaning high). Questions 25 to 30 were answered by text.
The questions address Quesenbery's usability dimensions~\cite{quesenbery2003dimensions} in the following manner: \textit{Effective} is addressed by questions 1, 19 and 21; \textit{Efficient} by question 17; \textit{Engaging} by questions 4, 13, 23 and 24; \textit{Error tolerant} by questions 2, 3 and 20; \textit{Easy to learn} questions 5-12, 14-18 and 22.
A pilot study was carried out to validate the process (see~\cite{DBLP:conf/quatic/CoutoRC14}).

\begin{table}[t]
\centering
\caption{The Questionnaire from stage 2.} \label{tab:quest}
  {\scriptsize
	\begin{tabular}{|l|l|l|l|}
	  \hline
  	  Q. nr. & \textbf{Question} & Q. nr. & \textbf{Question}\\
	  \hline
	  \hline 
	  1& Number of statements which required major &  16& Minutes spent in adjustments (to match RUS) \\
	  & changes in order to be mapped into RUS & & \\
	  \hline 
	  2& Number of statements which lost their meaning & 17& How close to NL is RUS \\
	  \hline 
	  3& Number of unsupported statements &  18& How easy was it to understand the tool\\
	  \hline 
	  4& How much sense does the user interface makes & 19& How this tool is preferred over VP\\
	  \hline 
	  5& How familiar was the terminology & 20& How useful and adequate was the output\\
	  \hline 
	  6& How much the tool helps in the specification & 21& How acceptable are the tools' limitations\\
	  \hline 
	  7& How easy to use is the RUS & 22& How easy to use is the tool \\
	  \hline 
	  8& How easily the participants understand the RUS & 23& How much the user liked the language \\
	  \hline 
	  9& How much easier is RUS to understand than NL & 24& How much the user liked the tool \\
	  \hline 
	  10& How easy is it to manipulate RUST & 25 & What became harder by using RUS (over NL)? \\
	  \hline 
	  11& Is NL easier to use than RUS & 26 & What became easier by using RUS (over NL)? \\
	  \hline 
	  12& Is NL easier to understand than RUS & 27 & What did you like in the language? \\	  
	  \hline
	  13 & Likelihood to adopt RUS & 28 & What did you dislike in the language? \\
	  \hline
	  14& How easy is it to understand RUST & 29 & What did you like in the tool? \\
	  \hline
	  15& How clear is the language & 30 & What did you dislike in the tool? \\
	  \hline
	\end{tabular}}
\end{table}

For the study proper, participants were distributed by four sessions. Each session took about 130 minutes: 30 minutes for presenting the tool and the language, 90 minutes for the study, 10 minutes for the questionnaire. 
In each session, the participants were gathered in a room, and asked to perform the scripts individually. 
No time limits were imposed. 
In each session the participants performed both the first and the second stages of the study in sequence.  

The scripts for first stage contained three tasks regarding the specification and interpretation of use cases. 
In \textbf{Task 1} the participants were asked to interpret and textually describe the \ac{RUS} use cases we had produced. In \textbf{Task 2} we handed the original use cases to their authors, which then evaluated how faithful to the original the textual descriptions made by the other participants were (i.e., how well they described the original specification). In \textbf{Task 3} the participants  were asked to compare their textual descriptions of the RUS use cases with the original use cases, and to point out any missing information from the \ac{RUS} version we had produced. 

The second stage was composed of four tasks. In \textbf{Task 4} the participants were asked to translate a textual scenario to the tabular notation that their were used to (similar to Fowler's approach, and using natural language to describe the interactions). The scenarios concerned a web application context. Three different scenarios were used: ``Upload a model to a repository'', ``Download a model from a webpage'' and ``Register on a group on a web application''. 
The scenarios were written in Portuguese as that was the participants' native language. The scenarios were translated into english for this paper. As an example the latter scenario was as follows:
\begin{quote}
\textit{A user clicks in ``groups'' link. The system shows the available groups. Next, the user views the list, and selects one to register. The user selects ``register'', the system registers the user in the group and shows a success message. If the group is private, after selecting ``register'' the system sends a message to the group author (with an admission request) and shows an information message, instead of success message.}
\end{quote}
In \textbf{Task 5}, the participants converted the previous use case into \ac{RUS}, using the \ac{uCat} tool. In \textbf{Task 6} they were asked to write another scenario directly in \ac{RUS} using \ac{uCat}. In \textbf{Task 7} participants handed the use cases to other participants, which interpreted them. Next, the descriptions where handed to the original authors which evaluated them.
At the end, the questionnaire was applied.

In the third stage we started by presenting both the \ac{RUS} language and the \ac{uCat} tool (during approximately 20 and 10 minutes, respectively). We demonstrated how the tool supports the language, by specifying a use case in the tool (adding a product to the shopping cart). 
Participants had the opportunity to interact with \ac{uCat} and to create themselves a specification, in order to get used to the tool. This took about 10 minutes. Note that, given their background, all participants were well versed in using software modeling tools. 
For this stage participants had to carry out a single task:  \textbf{Task 8}. They were asked to specify a set of four textual descriptions of usage scenarios for an \ec{} website in \ac{RUS}. 
Namely: 1) searching for a specific item by keyword in a website; 2) listing products from a specific category and accessing the details of a particular product; 3) performing registration and sign in; and, finally, 4) checking the recently viewed items. 
As an example, the first scenario was:
\begin{quote}
\textit{A user clicks in the "search'' link in the website. Then, the system shows a field where the user should insert the keyword to search, as well as the search criteria (price, date, etc.). When the user clicks ``ok'', the system performs a search (based in the given criteria), creates a result list and shows such list to the user. Finally, the user checks the resulting list.}
\end{quote}
This task took about 50 minutes to complete. Again no time restrictions were imposed. 
After completing the specification, participants answered the \ac{SUS} questionnaire~\cite{brooke1996sus}. The specifications were kept to validate the requirement patterns (as future work).
To perform the study, we gathered all the participants in the same room. Each participant had a laptop and was asked to install \ac{uCat}. It was handed a printed page containing all the usage scenarios and another containing the \ac{SUS} questionnaire. They were asked to perform the study individually, and at the end to save and send us the results (i.e., \ac{RUS} specifications and \ac{SUS}).

\subsection{Results of the experiment}

In all the three stages the participants demonstrated autonomy while performing the tasks. As presented next, in some steps some questions arose, but all regarding minor issues. 

Table~\ref{tab:res_p1} summarizes the results from the first stage of the study regarding a) whether the use case descriptions produced in Task 1 were correct (Task 2); and b) whether there were differences between the \ac{NL} and the \ac{RUS} use case descriptions (Task 3). In the second task, only 15 of the 18 participants answered the questions. 

In the second stage of the study, while writing the specifications in natural language was straightforward, writing them in the tool generated some questions. Most common questions regarded input mismatches (for instance, a trailing space in a statement, or how to write a multiple word entity such as ``work plan''). All the questions were easily answered, not affecting the study in a negative way.

An excerpt of a \ac{RUS} use case produced by a participant is shown in Table~\ref{tbl:ruc}, starting from a \ac{NL} use case. It corresponds to the success case  for the ``Register on a group on a web application'' scenario from Section~\ref{sec:setup}, and it is a case where the author of the \ac{NL} use case considered the RUS version a correct version of the original NL use case:
\textit{``The  description is in accordance with the specification''}.

\begin{figure}[tb]
\minipage{0.5\textwidth}
	\centering
	\includegraphics[width=0.95\textwidth]{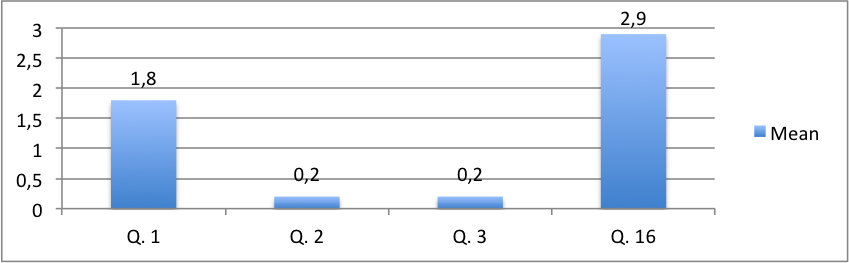}
	\caption{Questionnaire results for open questions.}
	\label{fig:resa}
\endminipage\hfill
\minipage{0.5\textwidth}
	\centering
	\includegraphics[width=0.95\textwidth]{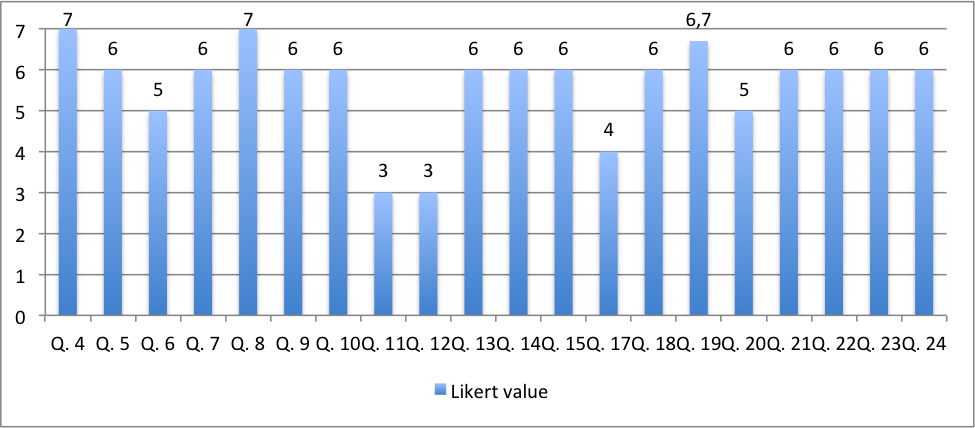}
	\caption{Questionnaire result for Likert scale questions.}
	\label{fig:resb}
\endminipage\hfill
\end{figure}

Figures \ref{fig:resa} and \ref{fig:resb} present the questionnaires results up to question 24 (mean value for open answers, and mode for Likert-scale answers, respectively).
These results will be discussed in Section \ref{sec:obj}.
The open questions in the questionnaire enabled participants to express their experience.
The first question (see Table~\ref{tab:quest}) was:
 \emph{``What became harder by using RUS (over NL)?''}. 
Three participants answered that nothing became harder, while eight referred the need to learn and adjust to the \ac{RUS} syntax. Six participants answered that it was to map more complex statements into \ac{RUS}.
On the contrary, when asked \emph{``What became easier by using RUS (over NL)?''}, seven participants answered that it was the interpretation, as the descriptions became  simpler. 
Four participants referred the standardization of the specification, while three referred that it became easier to create specifications. Three participants stated that it became easier to create specification (vs \ac{NL}), and one participant answered that it was easier to create correct specifications. 

Another asked question was: \emph{``What did you like in the language?''}. 
Thirteen  participants answered that it was its simplicity, two participants referred the standardization of the specifications, one the interpretation of use cases produced by other authors, another that it speeds up the specification process, and one referred nothing. To the question: \emph{``What did you dislike in the language?''}, nine participants answered that there was nothing that they disliked, three participants reported the need to adapt \ac{NL}, other three the limited set of keywords to support statements, a clear indicator that they did not understand \ac{RUST}. Other two participants reported the required learning time, and one participant how alternative scenarios could be specified (although that is an issue related more with the tool). 

We asked also if the users preferred the \ac{RUS} format (supported by \ac{uCat}), or another free text format input tool they knew. Ten participants stated that they preferred our approach because of the standardization provided by the format, six participants mentioned the lightweight interface, and two the easier way to specify alternatives. One user stated that it becomes closer to a programming process.
When asked the question: \emph{``What did you like in the tool?''}  most participants (seven), mentioned its simplicity, four the capability to validate the use cases while specifying them, three  the formatted input, and two  the representation of the information. One participant referred the familiar interface, and another the possibility for the tool to be a viable replacement for other tools. On the contrary, when asked:\emph{``What did you dislike in the tool?''} ten participants pointed nothing, three mentioned the restrictions on the specifications format, and three proposed improvements in the alternative scenarios specification. One referred minor issues (as bugs), and another stated that specifications became harder to read. 

\begin{table}
\parbox{.45\linewidth}{
\centering
\caption{RUS evaluation.}\label{tab:res_p1}

\begin{tabular}{|l|l|l|l|}
\hline
\textbf{Question} & $\surd$ & $\sim$ & $\times$ \\ \hline
\begin{tabular}[c]{@{}l@{}}a) Use case description \\  is correct\end{tabular} & 12      & 2          & 1         \\ \hline
b) Versions are identical                                                        & 14      & 4          & 0         \\ \hline
\multicolumn{4}{l}{\scriptsize($\surd$ -- correct; $\sim$ -- had issues; $\times$ -- incorrect)} 
\end{tabular}}~
\parbox{.45\linewidth}{\centering
\caption{Excerpt of use case made by a participant.}\label{tbl:ruc}
{\scriptsize\tt
\begin{tabular}{|l|l|l|}
\hline
1 & user clicks groups  & \\
2 &                  & system provides the groups \\ 
3 & user selects group  &                                 \\ 
4 & user clicks register &                                \\ 
5 &                   & system register user            \\ 
6 &                    & system provides  \\ 
  &                    & success\_message \\ \hline
\end{tabular}}}
\end{table}

Moving on to the third stage of the study, only sporadic questions were made regarding either syntactical details of the language (e.g. ``\textit{should I use a verb here?}'', or, ``\textit{should I split this specification in two statements?}''), how to perform some action in the tool (e.g. ``\textit{how do I save the specification?}''), or minor questions related with the scenarios e.g. ``\textit{should I specify the action of viewing an item?}''). We successfully answered all of the questions and the participants were able to proceed with the study.
In general they appreciated the fact that uCat was a lighter tool, and expressed that they would prefer this tool over the previously used one.

As with the first and second stages, we have also analyzed the specifications made by the participants. 
It could be seen that test subjects correctly used the syntax to create meaningful specifications.  
The \ac{RUS} specification  produced by one of the participants for the web search scenario is shown in Table \ref{tab:rus}. 
\begin{table}[htbp]
\caption{RUS use case for performing a search in a e-commerce website.}
\centering
{\scriptsize
\tt
\begin{tabular}{|c|l|l|}
\hline
1 & user clicks in search &\\
2 & & system shows the search\_field\\
3 & user inserts the keyword, search\_criteria &\\
4 & & system performs search\\
5 & & system creates list\\
6 & & system shows list\\
7 & user checks list &\\
\hline
\end{tabular}}
\label{tab:rus}
\end{table}

Regarding the specifications, 14 participants were able to correctly describe the use cases. From the 14 participants, 4 pointed out some issue with the description (not deterrent for the study itself).
Regarding the 4 that had issues, two participants pointed out missing descriptions of the alternative scenarios (despite their being in the specification), one pointed out a missing bit of context information (it was not stated that the group could be private, according to the aforementioned scenario), and another misunderstood ``information'' for ``success message''. Hence, the participants overall produced correct use case specifications for the given scenario. 

The goal of applying the \ac{SUS} questionnaire was to evaluate the usability of the tools' user interface.
It consists of ten questions about the system which are answered using a 5 point Likert scale (from 1 -- Strongly disagree -- to 5 -- Strongly agree).
From the answers provided, a score is calculated which has been shown to have a strong correlation with the perceived usability of the user interface being analyzed. 
\ac{uCat} scored a value of 74, meaning that it  has higher perceived usability than (approximately) 72\% of all products tested~\cite{brooke2013sus}. The corresponding grade is \textbf{B}.

\section{Discussion}
\label{sect:disc}

In this section we discuss the results of the participants for the proposed tasks. This give us useful indications about the language and the tool performance in a practical scenario. We discuss also the study results regarding the initial proposed objectives, and possible threats to validity.

\subsection{Analysis of the performed tasks}
Regarding the performed tasks, both the participant's results and observed behaviors provide us relevant insights. 
\textbf{Tasks 1 and 2} allowed us to conclude that participants were able to correctly understand the presented \ac{RUS}.
Only two of the descriptions produced by the participants were considered to have minor issues.
One had resulted from an interpretation error, where a use case element (\texttt{user}) was used instead of another (\texttt{player}). The other consisted on a poor description of the scenario. The participants described the steps themselves (leading to a  hard to interpret description).
On a positive note, none of the interpretation errors was due to either the language or the translation process.

The results from \textbf{Task 3} show that the participants were able to both understand and express the use cases in \ac{RUS} without issues, and that the language had enough expressiveness. None of the users reported missing information, which would lead to changes in the \ac{RUS} use case descriptions.
In fact, only five participants referred some missing detail; one stated that the \ac{RUS} version was more compact, and the remaining stated that both descriptions had the same information. 

From \textbf{Tasks 4, 5 and 6} we can draw several conclusions. The participants required an average of 27 seconds per statement (s/s) when creating the tabular \ac{NL} use case descriptions. 
While writing the same description in \ac{RUS}, the users required an average of 70s/s. That is somehow understandable considering that the users had no previous training, therefore needed an adaptation period. However, when writing a description for the second time,  the participants required an average of 52s/s (18s/s less), corresponding to an improvement of about 25\%.  
We applied the two-sample T-test (for iteration 1 and 2), for the null hypothesis that there is no difference between the two populations means. 
The result, for a confidence level of 95\%, states that the mean value of iteration 1  is greater. 
This clearly indicates that through practice, the users were able to reduce the time required to write \ac{RUS} statements.

In \textbf{Task 7} participants  were able to correctly describe the use cases written in \ac{RUS}. From the 18 descriptions, just in two were missing details pointed out. 
These were not related with the language description, rather with brevity of the users descriptions. This is another positive indication about the expressiveness of \ac{RUS}.  

The presented results, plus the participants' feedback, clearly indicate that they were receptive to our approach. Participants were able to perform the tasks independently.
The formalization of the use cases was possible with minor overhead, and there was no loss of information in the formalization process. 
Only one of the specifications produced in the second stage contained issues. In that case, the participant clearly did not understand the purpose of the language, as the remaining specifications were correct.
Analysis of the specifications from the third stage shows that, overall, participants made a correct usage of  \ac{RUS} syntax. All the specifications were valid, and correctly described the corresponding scenarios. 
The lack of issues and the participants' autonomy are good indicators regarding both the tool and language. 

The specifications varied in the used verbs, the number of statements, and in the names given to entities (e.g. the terms \textit{search\_link} and \textit{searchLink} were used by different participants to refer to a search link).
For instance, to describe the ``Download a model from a webpage'' scenario, (from stage 2), 7  participants used 7 or 8 statements, while the other 3 used 6, 12 and 13. 
Each scenario presented from \textbf{Task 8}, varied in the number of statements required by different participants to describe them in \ac{RUS}. Namely, scenario \textbf{1)} and \textbf{2)} varied from 5 to 9 statements, scenario \textbf{3)} from 7 to 12 statements and scenario \textbf{3)} from 3 to 9 statements.
These results demonstrate the existence of variability in the descriptions.

\subsection{Analysis of the study objectives} \label{sec:obj}

The previous section provided a general discussion of the results of the study.
Next we describe, in more detail, how each objective from Section \ref{sect:val_ap} relates to the results of the questionnaire from stage 2.

\textbf{Objective 1} is related to questions 9, 11, 12 and 16. The answers show that participants preferred the \ac{RUS} approach over \ac{NL}. 
Several factors contributed to this result (as presented next), but overall the participants liked the standardization provided by the language.

\textbf{Objective 2} is related to questions 1, 2, 3 and 22. On questions 1, 2 and 3, on average 1.8 statements (for an average of 14.2 statements) required some kind of adjustment when mapping into \ac{RUS}.
This result is a good indicator that the language has a good expressiveness. 
From question 22, we concluded that the participants expressed empathy with the language. The participants considered also that the \ac{RUS} language is easy to learn and \ac{RUST} is easy to understand and manipulate (mode 6, on questions 8 and 14 and 10 respectively), which contributes to achieve this objective. These answers are in line with the results from the tasks, previously presented.

\textbf{Objective 3} is related to questions 7 to 15. From these questions, it is possible to concluded that the participants considered the tool to be easy to understand and use, even when compared with \ac{NL}.

\textbf{Objective 4} is related to questions 5, 17, 19, 21 and 23. With a mode of 6 for these questions, the tool had a good acceptance by the participants (being easy to use and learn).
Question 23 provides also feedback for the tool, and follow the trend of the other questions.

\textbf{Objective 5}  is related to questions 4, 18 and 20.
These questions have modes of 6 and 7. The results show that participants are highly receptive of using our tool as replacement/complement to other tools. 

\textbf{Objective 6} is related to questions 6, 19 and 21, which have modes of 5 and 6. This result shows that the participants consider the tool able to support the language.

\textbf{Objective 7} was measured by the results provided by the \ac{SUS} questionnaire.  
A score of 74 in \ac{SUS} was very satisfying since this was the first formal usability test performed with \ac{uCat}. Furthermore, it is in accordance with previous results, which suggested that the tool interface is adequate \cite{DBLP:conf/quatic/CoutoRC14}.

Beyond the questionnaire, direct observation during the study enabled us to conclude that the tool played a relevant role in the specification process. First, by ensuring the correctness of the specifications, as the tool forces the participants to input valid \ac{RUS} statements, since only valid specifications are accepted. Second, the tool provided runtime feedback regarding the statements: once the user finished writing a statement, it was immediately verified and highlight if incorrect. Thus, the test subjects not only knew if the specification was valid, but what statements were invalid.

\subsection{Threats to validity}

While we feel the results provide positive feedback on the tool, a number of aspects must be taken into consideration.
With 18 and 21 test subjects in each group we feel we have a reasonable level of confidence in the results. 
Naturally, increasing the sample size would result in more reliable results.
Note however that \ac{SUS} in particular is known to provide good results for small sample sizes (5 test subjects is usually considered an acceptable number for early stage evaluations, according to \cite{lewis1994sample}). 

The participants' background is  a  difficult to address issue. 
Ideally, we should have a more diverse collection of test subjects. However, that was not easy to achieve.
The fact that a considerable number of participants did not have professional experience, might have affected, for example, their willingness to accept new tools as the time invested in the tools they currently use and the cost of adopting new tools is not overly large.
On the contrary, the diversity of computers used by the participants (e.g. operating systems) might have affected the time measurements performed on some tasks.  Ideally all the participant should have a similar setup, however that was beyond our control.

Still on the topic of the variability of conditions, and despite our efforts to avoid it, different sessions were performed at different hours. While some groups performed the tasks in the morning, other performed them the end of the afternoon. The effects on the study are arguable, but since some tasks required focus, the fatigue of the participants could have affected them. 

Finally, it can be argued that the scenario descriptions provided were too detailed, and close to the use cases language, making it easy to perform the translation from natural language to use cases. However, our focus at this stage was on use case specification, not requirements analysis, and in order for the study to have a viable size, we decided to write simple and small statements that the all users could translate in reasonable time. We introduced however, as much as possible, subjective phrases to give some room for variations in the specifications. Indeed, for the same specification, different participants presented specifications with a different number of lines.

\section{Conclusions and future work}
\label{sect:conc}

Our work focusses on the formalization of requirements.
We have developed a language (RUS) to express Use Case specifications and a tool (uCat) to support the language and the process of generating a knowledge base from the specifications. 
The formalization of the requirements is relevant as a means to provide a better representation for the information they contain. 
As with other approaches (e.g. Gherkin), this formalization opens the possibility to apply automated techniques. For instance, the possibility to generate test cases, or formally validate the requirements. In this paper we have presented the validation of the approach to formalize use case descriptions, as well as  of \ac{uCat} tool, which supports the process. 

In order to evaluate the language and its tool support, we performed a three stage study. The first two stages were performed with 18 participants where they interpreted, read, and created \ac{RUS} specifications. At the end they answered a set of questions regarding both the language and the tool. 
In the third stage, 21 participants were asked to interpret and create \ac{RUS} specifications for 4 scenarios using uCat. At the end they answered a \ac{SUS} questionnaire. 
Regarding the language, the participants successfully both interpreted it and created new specifications, even without previous training. Not only were they able to produce specifications in an acceptable time interval, but practice further reduced the time required to write them. 
The tool's performance results are also positive. It performed well and generated positive feedback, successfully supporting all the tasks in the study. 
As a whole, results indicate that the proposed use cases formalization approach is feasible, and that the tool provides good support for the approach. It is possible to write use cases, without a major effort and without losing expressiveness, such that they can be automatically formalized into the knowledge based. 

Performing the validation study was not only relevant to validate the formalization approach, but also the supporting tool. It is not only relevant to have tools which support formalization mechanisms, but it is also important for them to have a good acceptance by the final users. By successfully validating our tool, we are improving the changes for it to be adopted by the users, and ultimately foster the adaptation of formalization techniques.
Results indicate \ac{uCat} is an adequate tool to provide support for the specification process required to support the approach.
The tool performed well in supporting the specification process and scored a value of 74 in \ac{SUS}, with a corresponding grade of~B. 

Now that we have a viable approach to formally express use case specifications, and we are able to represent the knowledge continued in the specifications in a tractable manner, a number of venues of future work can be explored. Using verification techniques we can explore how to analyze the quality and correctness of the use case specification prior to development. For example, we might explore the consistency between different requirements.  Next we can explore how to use the use cases information to generate user interface prototypes for early requirements validation. These prototypes will allow easier validation of the proposed solution with users prior to actual development, thus reducing the likelihood of problems in the requirements analysis phase. Going a step further, we are exploring how to map requirements into architectural solutions, in order to speed up development. By mapping requirements patterns, which can be identified in the specifications, to architectural solutions that address them, it become possible to quickly generate an initial architectural solution for the system. Finally, we intend to generate test cases from the use case specifications in order to test the developed systems against their requirements.
The acceptance of both the approach and tool by the industry is yet to be validated, despiste the positive results attained so far. This kind of validation is relevant, in order to evaluate the performance of the tool in real world scenarios. Hence, it is left for future work to perform a validation study resorting to use cases, used in already implemented applications. 
Another aspect to address is the scalability of the tool, and test how it reacts to larger requirement specifications. Finally, at the moment the supporting tool is implemented as a standalone application. Integrating \ac{uCat} (as a plugin) on widely adopted tools (e.g. IBM Rational DOORS), would help to validate its adoption.

\vspace{-0.5em}
\bibliographystyle{eptcs}
\bibliography{sigproc}
\end{document}